\documentclass[11pt]{article} 
\usepackage{amsmath}
\usepackage{amssymb}
\usepackage{graphicx}
\usepackage{psfrag}
\usepackage{epsfig}
\begin{document}

\def \bsigma {\mbox {\boldmath $\sigma$}}

\title{\bf A layered neural network with three-state neurons optimizing the
mutual information}
\author{D. Boll\'e$^a$, 
        R. Erichsen Jr.$^b$, 
	and W.K. Theumann$^b$}
\maketitle
\begin{center}
$^a$Instituut voor Theoretische Fysica, Katholieke 
Universiteit Leuven, \\
Celestijnenlaan 200 D, B-3001 Leuven, Belgium\\
$^b$Instituto de F{\'\i}sica, Universidade Federal do Rio
Grande do Sul, \\ Caixa Postal 15051. 91501-970 Porto Alegre, RS, Brazil
\end{center}

\begin{abstract}
The time evolution of an exactly solvable layered feedforward neural
network with three-state neurons and optimizing the mutual information 
is studied for arbitrary synaptic noise (temperature). Detailed stationary
temperature-capacity and  capacity-activity phase diagrams are
obtained.  The model exhibits pattern retrieval, pattern-fluctuation
retrieval and spin-glass phases. It is found that there
is an improved performance in the form of both a larger critical capacity
and information content compared with three-state Ising-type layered 
network models. Flow diagrams reveal that
saddle-point solutions associated with fluctuation overlaps slow down 
considerably the flow of the network states towards the stable 
fixed-points.

\end{abstract}

\section{Introduction}
By now it is common knowledge that layered feedforward models
are the workhorses in practical applications of neural networks and,
hence, progress in the theoretical understanding of their 
capabilities and limitations should thus be welcome. Recently, it has been
shown  \cite{DK00,BV02} how information theory can be used to construct 
neural network models leading to optimal performance. Optimal for the
task of retrieving an embedded pattern when the network starts far from
it with a vanishingly small initial mutual information. For two-state 
networks this approach recovers the well-known Hopfield model \cite{H82}
and for 
three-state networks a Hamiltonian is found reminiscent of the  
Blume-Emery-Griffiths (BEG) model \cite{BEG71} (see \cite{ACN00} for  
further references in a spin-glass context) with a novel Hebbian-like
learning rule. 

Both the extremely diluted asymmetric version \cite{DK00,DK02,BDE02} and 
the fully connected version \cite{BV02,BV03,BBS02,BBSV02}  of this model 
have already been studied. These studies  reveal that the retrieval 
performance of these so-called BEG networks, 
compared with the one of other three-state networks of the same
architecture (see \cite{Y89,BS94} and references therein), is better in the
sense that there is a selective
increase in the information content of the network and that a considerably
larger retrieval region exists in the phase diagrams leading also to a
sizable increase in the critical capacity. In particular, new information
carrying states, the so-called quadrupolar or pattern-fluctuation retrieval
states appear. They play an explicit role in enhancing the retrieval
performance of the network and they might also be important in practical
applications. In pattern recognition, e.g., looking at a black and white
picture on a grey background, these states would describe the situations
where the exact location of the picture with respect to the background 
is known but, the details of the picture itself are not focused. 
Furthermore, these pattern-fluctuation retrieval states might be helpful 
in modelling these focusing problems discussed in the framework of
cognitive neuroscience \cite{SNK98}. Consequently, a study of
this BEG-model with a layered architecture is relevant.

Moreover, the study of this system is interesting in itself as an
exactly solvable non-trivial dynamical system. Compared with the extremely 
diluted 
asymmetric architecture it contains correlations among the neurons because
of the presence of common ancestors, although there are no feedback
loops \cite{BKS96}. Nevertheless, these correlations can 
be handled exactly giving rise to layer-to-layer evolution equations in
closed form, as also seen in Ising-type models \cite{DKM89,BSV94}. Since 
in the diluted BEG-architecture long transients appear due to the presence
of saddle-point solutions which slow down considerably the dynamics of the
network \cite{BDE02}, it is worthwhile to find out whether such a dynamical behavior
survives in the presence of correlations.
Finally, one also likes to find out what further differences there exist
between these  BEG-architectures and their analogues in other 
three-state models.

The outline of the paper is the following. In Section 2 we introduce
the three-state layered BEG network model and the relevant
macroscopic variables. In Section 3 we solve the dynamics of this model
by deriving the recursion relations for these variables.  We  discuss
the results in Section 4, for both the stationary phase diagrams and 
the dynamic flow diagrams. We end with some concluding remarks in 
Section 5.

\section{The Model}

Consider a network that consists
of $L$ layers, where each layer index may be taken as a time step $t$.
On each layer there are $N$ neurons that can take values $\sigma_i^{t}$,
$t=1,...,L; i=1,...,N$ from the set ${ S}\equiv\{-1,0,+1\}$, where
$\pm 1$ denote the active states. A macroscopic number of $p=\alpha N$
ternary patterns is taken from a set of independent identically
distributed random variables $\{\xi^{{\mu},t}_i=0,\pm 1\}$,
$\mu=1,...,p$, where $\pm 1$ are the active patterns, with the
following probability distribution on layer $t$,
\begin{equation}
\mbox{Prob}(\xi^{{\mu},t}_i)
   = a\delta(|\xi^{{\mu},t}_i|^{2}-1)+(1-a)\delta(\xi^{{\mu},t}_i) \, . 
 \label{1}
\end{equation}
This distribution is assumed to be the same for every layer and the mean
over it, $a=\overline{(\xi^{{\mu},t}_i)^2}$,  denotes the 
activity of the patterns. Together with this, a set $\{\eta^{{\mu},t}_i\}$ 
of normalized fluctuations of the binary patterns $(\xi^{{\mu},t}_i)^2$ 
about their average is introduced,
\begin{equation}
  \eta^{{\mu},t}_i=((\xi^{{\mu},t}_i)^{2}-a)/a(1-a) \,.
  \label{2}
\end{equation}
Both, patterns and fluctuations, are embedded in the network by
means of a generalized learning rule that consists of two Hebbian-like
parts,
\begin{equation}
J_{ij}^{t}= 
\frac{1}{a^2 N} \sum_{\mu=1}^{p} \xi^{{\mu},t+1}_i\xi^{{\mu},t}_j
 \,\,\,,   \quad 
K_{ij}^{t}= 
\frac{1}{N} \sum_{\mu=1}^{p} \eta^{{\mu},t+1}_i \eta^{{\mu},t}_j \,\,\,.
  \label{3}
\end{equation}
The first part is the usual rule in a three-state layered network
that codifies the patterns, while the second part codifies the
fluctuations of the binary active patterns $(\xi^{{\mu},t}_i)^{2}$
about their average.

Given the configuration on the first layer, $\bsigma_N^{1}\equiv
\{\sigma_j^{1}\}, j=1,...,N$, the state of a unit $\sigma_i^{t+1}$
on layer $t+1$ is determined by the configuration
$\bsigma_N^{t}$ of the units in the
previous layer according to the stochastic law
\begin{equation}
\mbox{Prob}(\sigma_i^{t+1}=s\in{ S}|\bsigma_N^{t})
  =\frac{\exp[-\beta\epsilon_i(s|\bsigma_N^{t})]} 
   {\sum_{s\in{S}}\exp[-\beta\epsilon_i(s|\bsigma_N^{t})]} \,\,\,. 
      \label{4}
\end{equation}
In this expression, the single-site energy function for unit $i$ on 
layer $t+1$, $\epsilon_i(s|\bsigma_N^{t})$ is given by
\begin{equation}
\epsilon_i(s|\bsigma_N^{t})
 =-sh_i^{t+1}(\bsigma_N^{t})-s^2\theta_i^{t+1}(\bsigma_N^{t}) \,\,\,, 
   \label{5}
\end{equation}
where
\begin{equation}
h_i^{t+1}(\bsigma_N^{t})=\sum_{j=1}^NJ_{ij}^{t}\sigma_j^{t},
  \quad
\theta_i^{t+1}(\bsigma_N^{t})=\sum_{j=1}^NK_{ij}^{t}{(\sigma_j^{t})}^2 \, 
 \label{6}
\end{equation}
are the local fields acting on that unit. In distinction to the
usual three-state model \cite{BSV94}, where the coefficient of the
quadratic part in $\epsilon_i(s|\bsigma_N^{t})$ is an externally
adjustable threshold parameter, we have here a random
self-adjusting function
$\theta_{i}^{t+1}(\{\sigma_j^{t}\},\{\eta^{{\mu},t}_k\})$ that
depends on both the states of the network and the patterns. 

Next, we consider the relevant quantities that describe the
performance of the network. For both, the macroscopic order
parameters and the mutual information, we need the conditional
probability distribution $\mbox{Prob}(\sigma_i^{t}|\xi^{\mu,t}_{i})$ that 
a neuron $i$ is in the state $\sigma_i^{t}$ on layer $t$ given that
the site $i$ of the stored pattern to be retrieved is
$\xi^{\mu,t}_{i}$. As a consequence of the independence of the
states of the units on a given layer, it is sufficient to
consider the distribution for a single typical neuron, so we can
omit the index $i$. We also omit the layer index $t$ and take from
previous work \cite{BD00}
\begin{equation}
\mbox{Prob}(\sigma|\xi^{\mu})=
    (s_{\xi}+m^{\mu}\xi^{\mu}\sigma)\delta(\sigma^{2}-1)+
                           (1-s_{\xi})\delta(\sigma),
    \label{7}
\end{equation}
where
\begin{equation}
      s_{\xi}= s^{\mu}+l^{\mu}(\xi^{\mu})^{2},\quad
      s^{\mu}={q_0-an^{\mu}\over 1-a},\quad
      l^{\mu}={n^{\mu}-q_0\over 1-a}.
 \label{8}
\end{equation}
Here, $m^{\mu}=\overline{\langle\sigma
\rangle_{\sigma|\xi}\xi^{\mu}/a}$ is the thermodynamic limit,
$N\rightarrow\infty$, of the retrieval overlap
\begin{equation}
m_{N}^{\mu}={1\over aN}\sum_{i}\sigma_{i}\xi_{i}^{\mu}
\label{9}
\end{equation}
between the state of the network and pattern $\{\xi_{i}^{\mu}\}$,
where the brackets denote the average over the probability 
distribution Eq.(7) and the bar denotes the configurational 
average over the patterns. The other parameters are the 
thermodynamic limits $q_0=\overline{\langle\sigma^{2}
\rangle_{\sigma|\xi}}$ and 
$n^{\mu}=\overline{\langle\sigma^{2}\rangle_{\sigma|\xi}
(\xi^{\mu})^{2}/a}$, of the neural (dynamical) activity, respectively the
activity overlap
\begin{equation}
(q_0)_N=
   {1\over N}\sum_{i}\sigma_{i}^{2}\,,
   \quad
 n^{\mu}_{N}={1\over aN}\sum_{i}\sigma_{i}^{2}(\xi_{i}^{\mu})^{2}\,.
\label{10}
\end{equation}
Finally, $l^{\mu}= \overline
{\langle\sigma^2\rangle_{\sigma|\xi}{\eta^{\mu}}}$, is the
thermodynamic limit of the fluctuation overlap between the
binary state variables $\sigma_{i}^2$ and $\eta_{i}^{\mu}$ defined
as,
\begin{equation}
l_{N}^{\mu}={1\over N}\sum_{i}\sigma_{i}^2\eta_{i}^{\mu}.
\label{12}
\end{equation}
As is clear from its definition, the fluctuation overlap is connected
with the activity overlap.
We remark that an underlying assumption that leads to the BEG model and 
that should be preserved in the implementation for any network
architecture is that the dynamic activity $q_{0}\sim a$. The
necessity of such an activity control system has been emphasized
before (cf. \cite{DB98,O96} and references therein).

Next, the mutual information between patterns and neurons, regarding the
patterns as the inputs and the neuron states as the output of the
network channel on each layer, is an architecture independent 
property given by \cite{Sh48,Bl90}
\begin{equation}
I^{\mu}(\sigma,\xi^{\mu})
    =S(\sigma)-\overline{S(\sigma|\xi^{\mu})},
\label{13}
\end{equation}
where
\begin{equation}
S(\sigma)=-q_0\ln(q_0/2)-(1-q_0)\ln(1-q_0)
\label{14}
\end{equation}\\
is the entropy and $\overline{S(\sigma|\xi^{\mu})}=
aS_{a}+(1-a)S_{1-a}$
is the  equivocation term with
\begin{eqnarray}
S_{a}&=&-c_{+}^{\mu}\ln c_{+}^{\mu}-c_{-}^{\mu}\ln c_{-}^{\mu}
-(1-n^{\mu})\ln(1-n^{\mu})\nonumber\\
S_{1-a}&=&-s^{\mu}\ln(s^{\mu}/2)-(1-s^{\mu})\ln(1-s^{\mu}).
\label{15}
\end{eqnarray}
Here, $c_{\pm}^{\mu}=(n^{\mu} \pm m^{\mu})/2$ and $s^{\mu}$ is the
parameter in the conditional probability $\mbox{Prob}(\sigma|\xi^{\mu})$.
The mutual information can then be used to obtain the information
$i^{\mu}=I^{\mu}\alpha$, where $\alpha=p/N$ is the storage ratio
of the network.

\section{Dynamics: Recurrence Relations}

To solve the dynamics and obtain the recurrence relations for the 
macroscopic variables we
need the expressions for the local fields which can be written as
\begin{equation}
h_i^{t+1}=\frac{1}{a}\sum_{\mu}\xi^{{\mu},t+1}_im^{{\mu},t}\,\,\,,
   \quad
\theta_i^{t+1}=\sum_{\mu}\eta^{{\mu},t+1}_il^{{\mu},t}\,\,\,,
\label{16}
\end{equation}
in terms of the actual overlaps
\begin{equation}
m^{{\mu},t}=\overline{\langle\sigma^{t}\rangle \xi^{{\mu},t}/a}
    \,\,\,,\quad
l^{{\mu},t}=\overline{\langle(\sigma^{t})^{2}\rangle{\eta^{{\mu},t}}}
   \,\,\,. 
   \label{17}
\end{equation}
At this point we remark that the fluctuation overlap $l^{\mu,t}$ can be
viewed as the retrieval
overlap between the binary states $\{\sigma_{i}^{2}\}$ and the
patterns $\{\eta_{i}^{\mu,t}\}$ and it is, in general, independent 
of the retrieval overlap $m^{\mu,t}$. One would expect the fluctuation 
overlap to become relevant for larger synaptic noise when the 
states of the network no longer distinguish between the active 
patterns. Indeed, it can be finite in a state of dynamic activity 
without necessarily a finite retrieval overlap $m^{\mu,t}$, as has 
been found before for both the extremely diluted and the fully 
connected network. As will be seen in the next Section, the fluctuation
 overlap 
is responsible for an enhancement of the information in most of 
the retrieval regime and for a finite information carried in the 
absence of retrieval.

In Eq. (\ref{17}) the brackets denote thermal averages with the probability
distribution Eq. (4). Now, $\langle\sigma^{t}\rangle$ and
$\langle(\sigma^{t})^{2}\rangle$ are given, respectively, by
\begin{equation}
F_{\beta}(h^{t},\theta^{t})=
\frac{\sinh(\beta h^{t})}
          {\frac{1}{2}e^{-\beta\theta^{t}}+\cosh(\beta h^{t})}\,\,, 
\quad
G_{\beta}(h^{t},\theta^{t})=
\frac{\cosh(\beta h^{t})}
     {\frac{1}{2}e^{-\beta\theta^{t}}+\cosh(\beta h^{t})}\,\,\,
    \label{19}
\end{equation}
which, in the zero temperature limit, $\beta\rightarrow\infty$,
become
\begin{equation}
F_{\infty}=\mbox{sign}(h)\Theta(|h|+\theta)\,\,\,,\,\,\,\,\,
G_{\infty}=\Theta(|h|+\theta)\,\,\,, \label{20}
\end{equation}
where $\Theta(x)$ is the usual step function.

We assume that a single pattern, $\xi^{1,t}$, and fluctuation,
$\eta^{1,t}$, are  condensed at each layer, that is, $m^{1,t}$ 
and $l^{1,t}$ are of order $ O(1)$, and that $m^{{\mu},t}$ and 
$l^{{\mu},t}$, $\mu>1$, are
of order $O(1/\sqrt{N})$. We call the former $m^{t}$
and $l^{t}$, respectively. In accordance with this, we also assume
that in Eq. (8) $n^{t}=n^{\mu,t}$ and $s^{t}=s^{\mu,t}$ are
both of $O(1)$, for $\mu=1$, and we denote the information content
of interest $i=i^{1}$. 
Following \cite{DKM89} each local field may then be separated into
a signal term and a noise 
\begin{equation}
h_i^{t+1}=\frac{1}{a}\xi^{1,t+1}_im^{t}+z\Delta^{t}\,\,\,,
  \quad 
\theta_i^{t+1}=\eta^{1,t+1}_il^{t}+w\Omega^{t}\,\,\,, 
\label{21}
\end{equation}
where $z$ and $w$ are Gaussian random variables with zero mean and
unit variance. The layer-dependent variances of the local fields
are given by
\begin{equation}
(\Delta^{t})^{2}=\frac{1}{a}\sum_{\mu>1}(m^{{\mu},t})^{2}\,\,\,,
\quad
(\Omega^{t})^{2}=\frac{1}{a(1-a)}\sum_{\mu>1}(l^{{\mu},t})^{2}
\,. 
\label{22}
\end{equation}
Together with Eqs. (\ref{17})-(\ref{19}), we have thus recurrence
relations for the overlaps and for the variances of the local fields.

The recurrence relations for the overlaps and the connecting
equations for the other dynamical variables become
\begin{eqnarray}
\label{23}
m^{t+1}&=&\int Dz\int Dw \,\,
   F_{\beta}\left(\frac{m^{t}}{a}+z\Delta^{t}\,,\,
               \frac{l^{t}}{a}+w\Omega^{t}\right)   \\
\label{24}
n^{t+1}&=&\int Dz\int Dw \,\,
    G_{\beta}\left(\frac{m^{t}}{a}+z\Delta^{t}\,,\,
                \frac{l^{t}}{a}+w\Omega^{t}\right)   \\
\label{25}
s^{t+1}&=&\int Dz\int Dw \,\,
     G_{\beta}\left(z\Delta^{t}\,,\,
                    \frac{-l^{t}}{1-a}+w\Omega^{t}\right) \,\,\,,
\end{eqnarray}
where, as usual, $Dx=\exp(-x^{2}/2)dx/\sqrt{2\pi}$. These
expressions yield the fluctuation overlap
$l^{t+1}=n^{t+1}-s^{t+1}$ and the dynamic activity
$q_{0}^{t}=an^{t} + (1-a) s^{t}$. A further variable,
$q_{1}^{t}=\overline{{\langle\sigma^{t}\rangle}^{2}}$, is
introduced in the derivation of the recurrence relations for the
variances of the two noises \cite{DKM89} and it is given by
\begin{eqnarray}
q_{1}^{t}&=&\int Dz\int Dw \,\,
   \left[a\,F^{2}_{\beta}\left(\frac{m^{t}}{a}+z\Delta^{t}\,,
                    \frac{l^{t}}{a}+w\Omega^{t}\right) \right. 
		    \nonumber \\
        &&+ \left. (1-a)\,F^{2}_{\beta}\left(z\Delta^{t}\,,
               \frac{-l^{t}}{1-a}+w\Omega^{t}\right)\right] \,. 
  \label{26}
\end{eqnarray}
Introducing the susceptibilities with respect to the $h^t$ and the
$\theta^t$  fields
\begin{equation}
\chi^{t}=A\beta (q_{0}^{t}-q_{1}^{t}) \,,
\quad
\psi^{t}=B\beta (q_{0}^{t}-p_{1}^{t})\,\,\,, 
\label{28}
\end{equation}
where $A=1/a$, $B=1/a(1-a)$ and
\begin{eqnarray}
p_{1}^{t}&=&\int Dz\int Dw \,\,
    \left[a\,G^{2}_{\beta}\left(\frac{m^{t}}{a}+z\Delta^{t}\,,\,
                  \frac{l^{t}}{a}+w\Omega^{t}\right) \right.
		  \nonumber \\
        &&+ \left. (1-a)\,G^{2}_{\beta}\left(z\Delta^{t}\,\,,\,
            \frac{-l^{t}}{1-a}+w\Omega^{t}\right)\right] \,\,, 
\label{29}
\end{eqnarray}
the recurrence relations for the Gaussian noises become
\begin{equation}
(\Delta^{t+1})^{2}=\alpha A^2q_{0}^{t}
+(\chi^{t})^{2}(\Delta^{t})^{2} \,\,, \quad
(\Omega^{t+1})^{2}=\alpha B^{2}q_{0}^{t}
+(\psi^{t})^{2}(\Omega^{t})^{2}\,\,\,. 
   \label{31}
\end{equation}

In contrast to these equations, the variances of the two local
fields in the extremely diluted network do not depend on the
susceptibilities but are  simply given by
the $q_0^t$-terms  \cite{DK00}. Since one expects
somewhat different behavior for the two architectures, it may be
interesting to see how the properties of the layered network model
changeover to those of the extremely diluted network. This can be
achieved most easily by means of a single amplitude $D$, varying between
$1$ and $0$,
in front of both second terms in the variance of the noises, as we will
discuss at the end of the next Section. 

With the above equations we also get the time evolution of the
information $i$ by means of Eqs. (\ref{13}) to (\ref{15}). The recurrence
relations for the macroscopic order parameters and the connecting
equations can now be used to study the evolution of the network
and to determine the properties of the stable stationary states.
The stationary states are reached when $m\equiv m^{t+1}=m^{t}$,
$n\equiv n^{t+1}=n^{t}$ and $s\equiv s^{t+1}=s^{t}$. Then $l\equiv
l^{t+1}=l^{t}$ and also $q_{0}^{t}$, $q_{1}^{t}$ and $p_{1}^{t}$
reach stationary values.

\section{Thermodynamics and flow diagrams}

In this section we study both the stationary solutions and the flow
diagrams for the layered BEG-model. Concerning the stable stationary
states of the network we find three kinds of phases. We have one or 
more retrieval phases $R(m>0,l>0,q_{1}>0)$, one or more  
pattern-fluctuation retrieval phases $Q(m=0,l>0,q_{1}>0)$, and a 
spin-glass phase $SG(m=0,l=0,q_{1}>0)$.
All three are sustained activity solutions in the sense that $q_{0}>0$.

The existence of the $Q$ phases can be understood as follows. A
non-zero $l$ will appear when the active binary neuron states
$\sigma^{2}$, that do not distinguish between $\pm 1$ neurons,
coincide with the active patterns. At the same time the actual
$\pm 1$ active neuron states may fail to recognize the active
patterns, meaning $m=0$. This is expected to occur at high $T$
where a stable $Q$ phase should appear. There is a finite $q_{1}$
for either form of the active neuron states. The presence of a
stable $Q$ phase only at high $T$ has been checked already for both the
extremely diluted \cite{BDE02} and the fully connected network 
\cite{BV02}. In particular, it appears in the phase diagram for 
$\alpha=0$, which is independent of the architecture. 

Since for the extremely diluted network it is found
that the $Q$ phase is not stable at zero-temperature, but
is instead a saddle-point \cite{BDE02} even for non-zero $\alpha$,
we consider first in Fig. 1 the capacity-activity phase diagram at
$T=0$ for the layered network.

\begin{figure}[h]
\vspace*{0.5cm}
\center{
\rotatebox{270}{
\includegraphics[width=.46\textwidth,height=10cm]{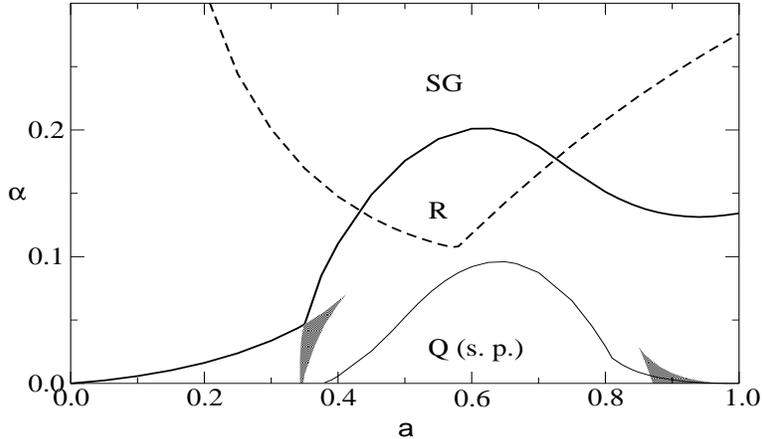}
}}
\caption{The capacity-activity $(\alpha-a)$ phase diagram for the BEG
and $Q=3$-Ising networks at $T=0$. The meaning of the regions and of
the lines is explained in the text.}
\end{figure}

There is a stable retrieval phase
below the heavy solid line and a second retrieval phase with a
smaller overlap appears in the lower shaded triangular regions.
The pattern-fluctuation retrieval states are only saddle-point 
solutions below the
light solid line. There is also everywhere a stable spin-glass
solution and all the lines denote discontinuous transitions. For
comparison, the heavy  dashed line shows the retrieval phase boundary 
for the optimal three-state Ising layered network, optimal in the sense 
that the adjustable threshold parameter $\theta$ was chosen to optimize
the storage capacity $\alpha$. Clearly, for intermediate activity 
$a\in (0.435,0.727)$ the BEG network has a larger
critical storage capacity than the Ising network \cite{BSV94}.

Another performance measure is the information content $i$ of the
network and in Figs. 2 we show the information-capacity diagrams
for various activities, at $T=0$, for the BEG and Ising networks.
Again, the BEG performance is better for intermediate activity and
the information content is purely due, at this temperature, to the only
stable retrieval phase. At larger activity, $a=0.8$ say,  
the BEG and Ising networks compete for better performance 
at intermediate or larger $\alpha$ values, as seen in Fig. 2c.

\begin{figure}
\center{
\rotatebox{270}{
\includegraphics[width=.35\textwidth,height=7cm]{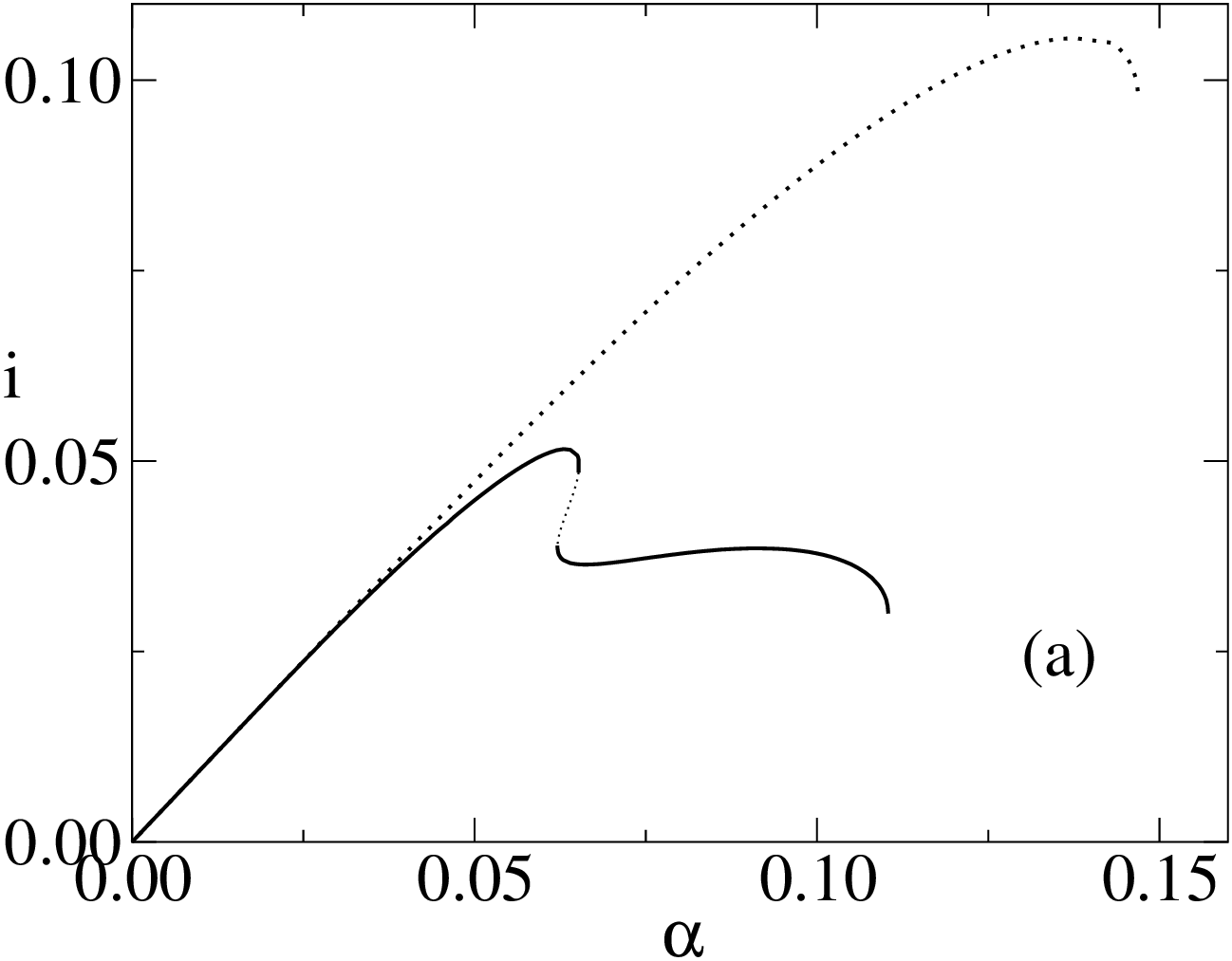}}}
\vspace*{0.5cm}
\center{
\rotatebox{270}{
\includegraphics[width=.35\textwidth,height=7cm]{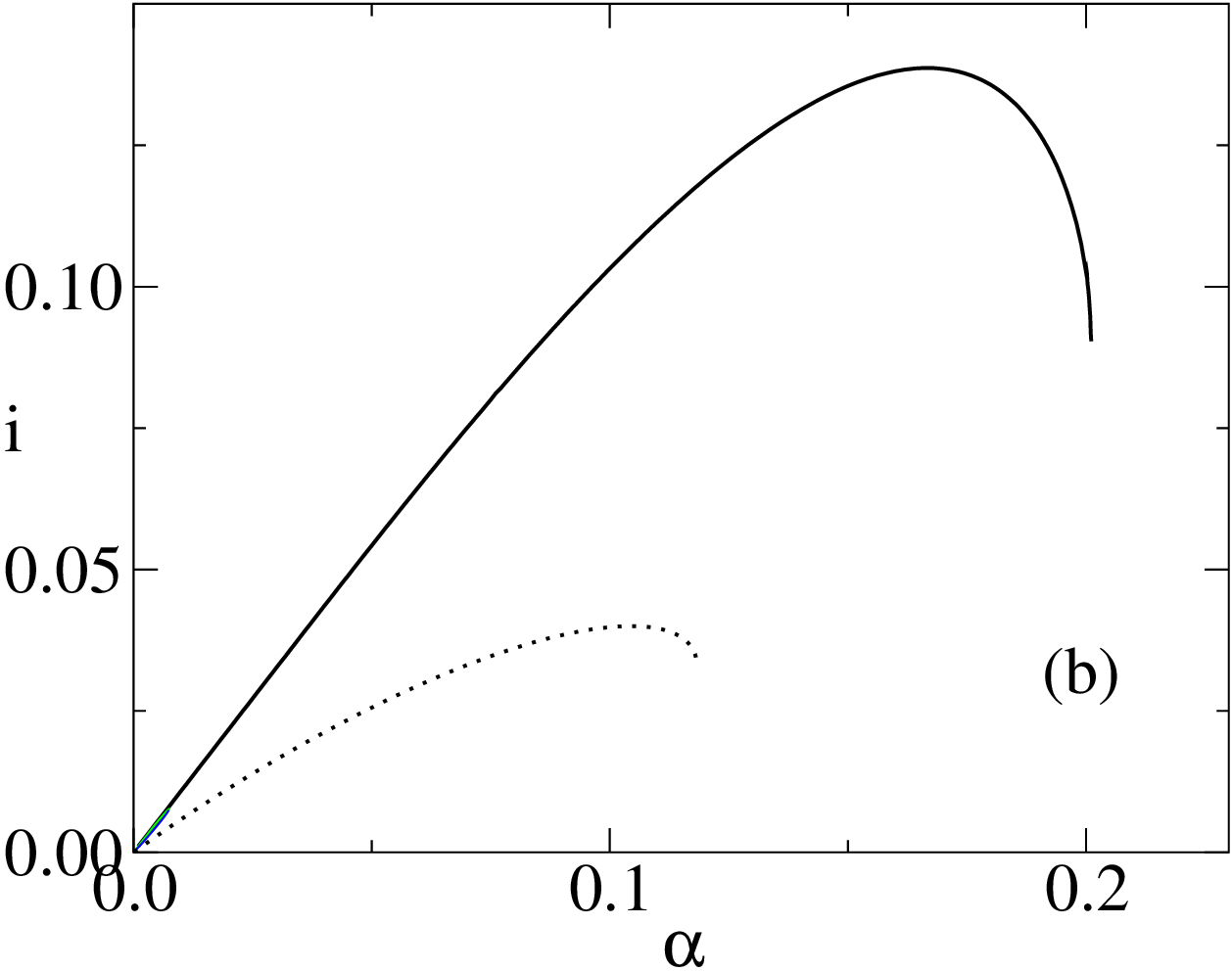}
}}
\vspace*{0.5cm}
\center{
\rotatebox{270}{
\includegraphics[width=.35\textwidth,height=7cm]{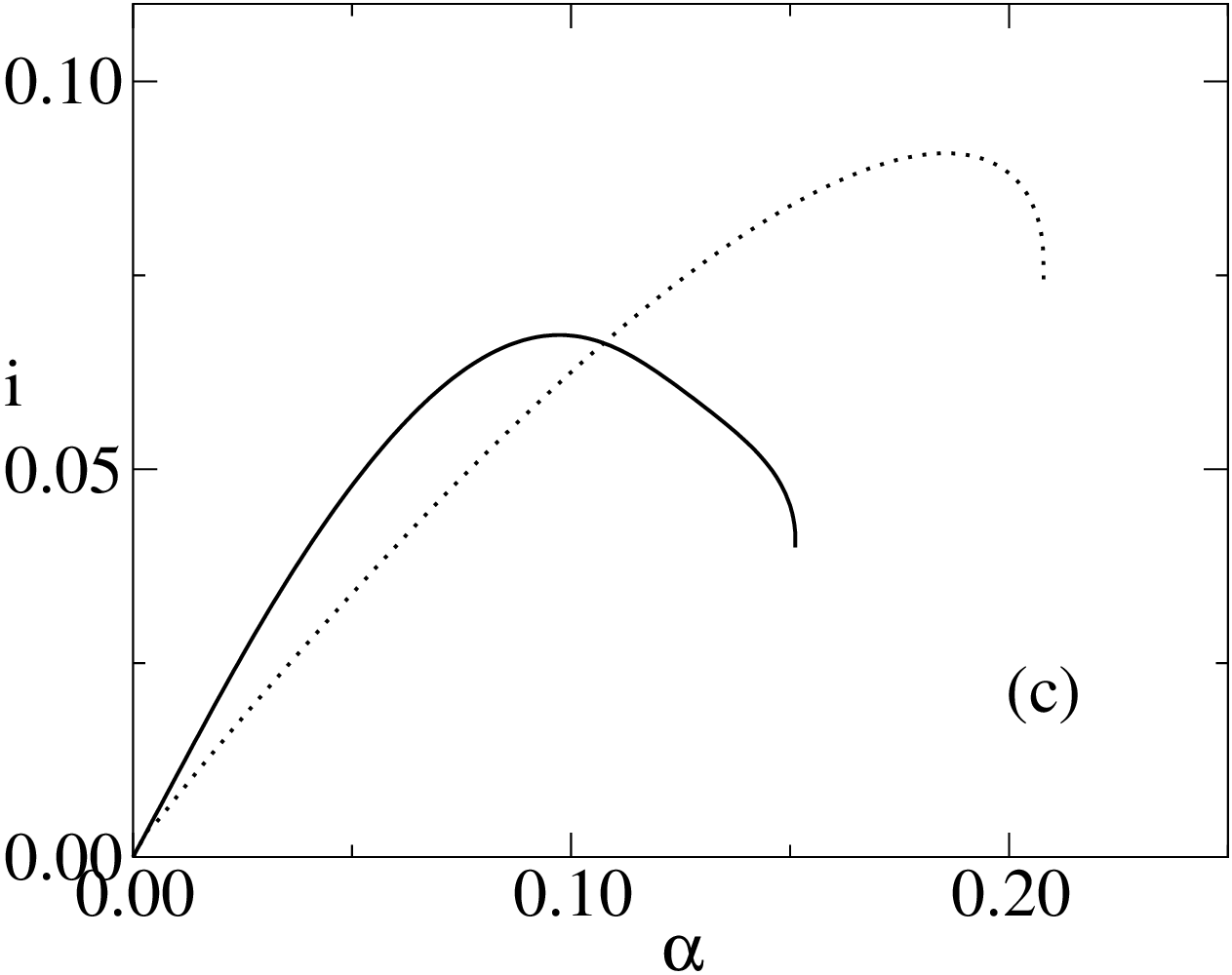}
}}
\caption{The mutual information $i$ as a function of the capacity
$\alpha$ for the BEG (solid line) and the Q=3-Ising (dotted line) 
networks for
$T=0$ and $a=0.4$ (a) $a=0.6$ (b) and $a=0.8$ (c).}
\end{figure}

In order to highlight the role of the temperature and the activity in
the appearance of a stable $Q$ phase, we consider the
temperature dependence for given activities and show in Fig. 3a
the temperature-capacity diagram for $a=0.4$ and in Fig. 3b for
$a=0.8$. In the first one there is only a stable retrieval phase
(a second one in the shaded area) below the heavy phase boundary,
in addition to the spin-glass phase. There are no $Q$ states, even 
not saddle-point solutions, in this region. In the case of $a=0.8$, 
instead, there is a single stable retrieval phase below the solid 
(dotted) heavy lines which denote discontinuous (continuous) 
transitions, respectively, with a tricritical point at $T\approx 0.753$ 
and $\alpha\approx 0.018$. There is also a stable (unstable
saddle-point) $Q$ phase above (below) the heavy lines, as
indicated in the figure, and the $Q$ phase solutions end at
discontinuous transitions shown as solid light lines. There is
also everywhere a stable spin-glass solution which only disappears
as $T\rightarrow\infty$. It is clear from Fig. 3b that a stable
$Q$ phase will only appear for sufficiently large synaptic
noise and large activity, a feature also found for the extremely
diluted \cite{BDE02} and the fully connected network \cite {BV02}.

\begin{figure}
\center{
\rotatebox{270}{
\includegraphics[width=.35\textwidth,height=7cm]{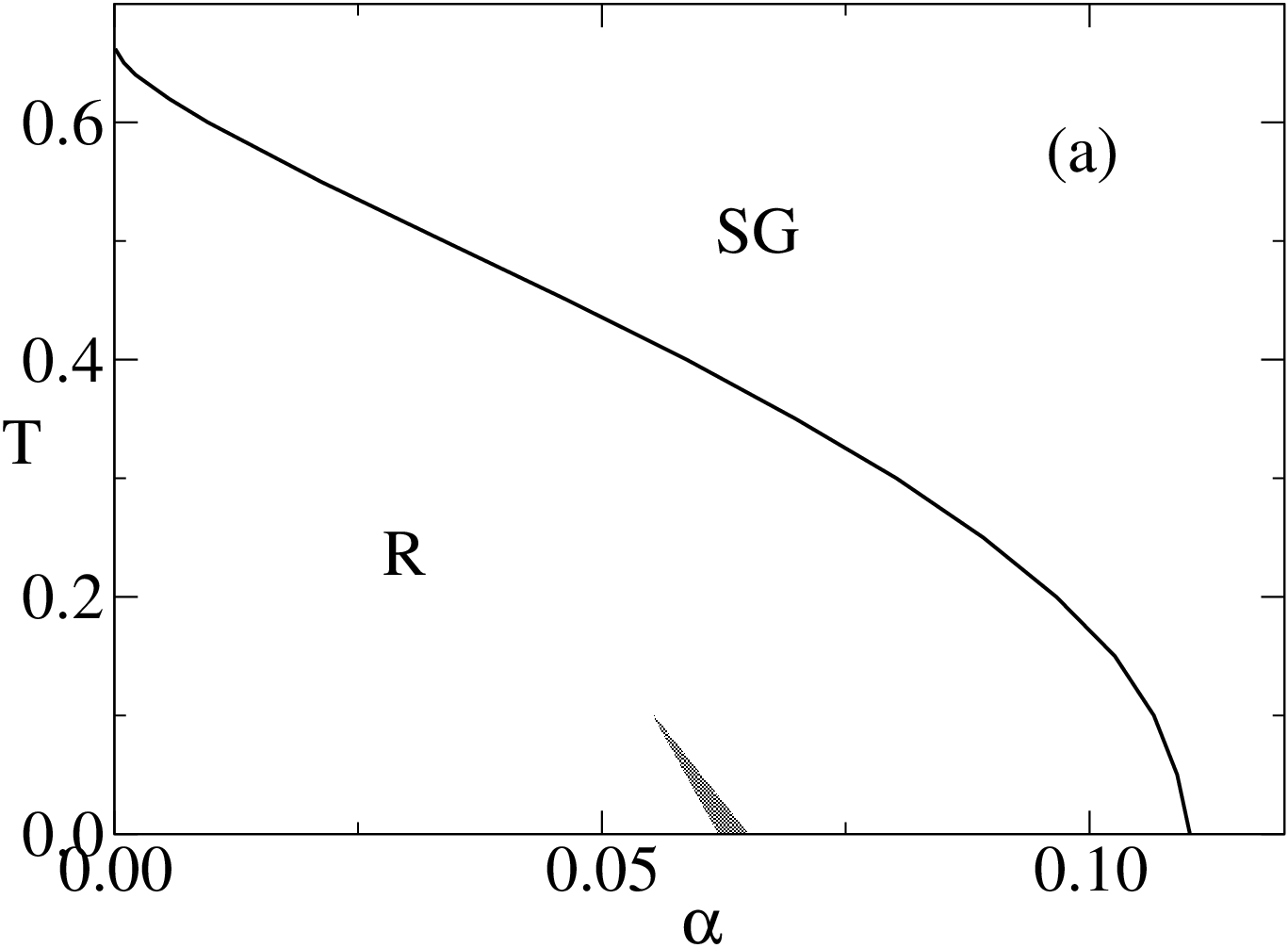}}}
\vspace*{0.5cm}
\center{
\rotatebox{270}{
\includegraphics[width=.35\textwidth,height=7cm]{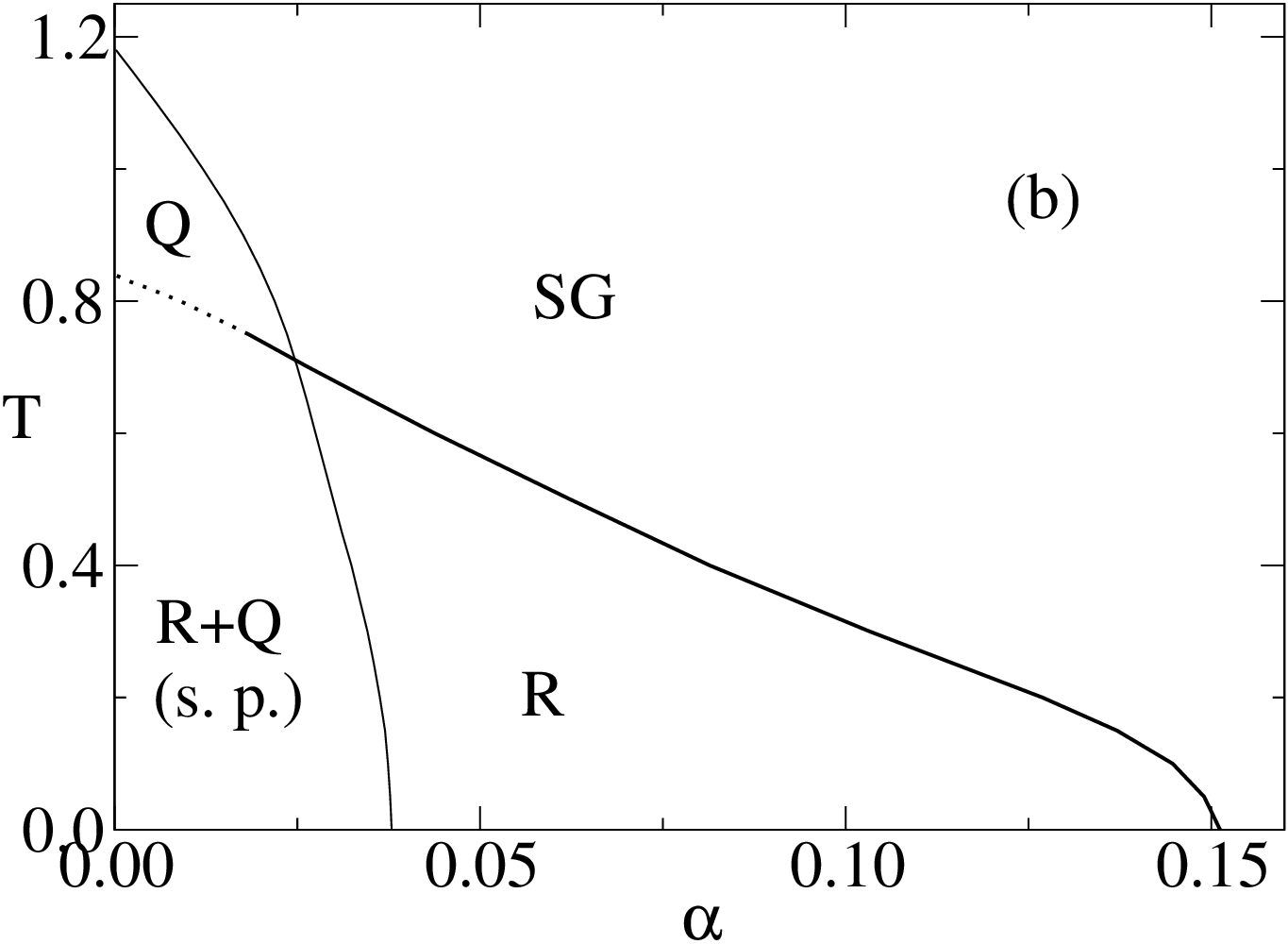}
}}
\caption{The temperature-capacity $(T,\alpha)$ phase diagram for the BEG
network for $a=0.4$ (a)  and $a=0.8$ (b).
The meaning of the regions and lines are explained in the text.}
\end{figure}

We consider now the role of the activity and show the capacity-activity
phase diagrams in Fig. 4a, for $T=0.4$, and in Fig. 4b, for $T=0.8$. 
In the case of the former there is a stable retrieval phase below 
the heavy solid line and a second stable retrieval phase in the 
shaded region. Again, there is a considerably enhanced storage 
capacity for retrieval, $\alpha\approx 0.119$, for an intermediate 
activity of $a\approx 0.676$. The pattern-fluctuation retrieval states
 are only 
saddle-point solutions and this is the case below the light solid 
line. When $T=0.8$ there is a stable retrieval phase below the 
heavy dotted (continuous transition) and heavy
solid (discontinuous transition) lines which merge at a tricritical
point at $a\approx 0.839$ and $\alpha\approx 0.0036$. The
pattern-fluctuation retrieval solution is a stable phase below the 
light solid line
and only a saddle point below the heavy and light dotted lines.
There is now a finite storage capacity, $\alpha\approx 0.022$, at
$a\approx 0.8$ for the retrieval of active patterns as a
pattern-fluctuation retrieval phase.

\begin{figure}
\center{
\rotatebox{270}{
\includegraphics[width=.35\textwidth,height=7cm]{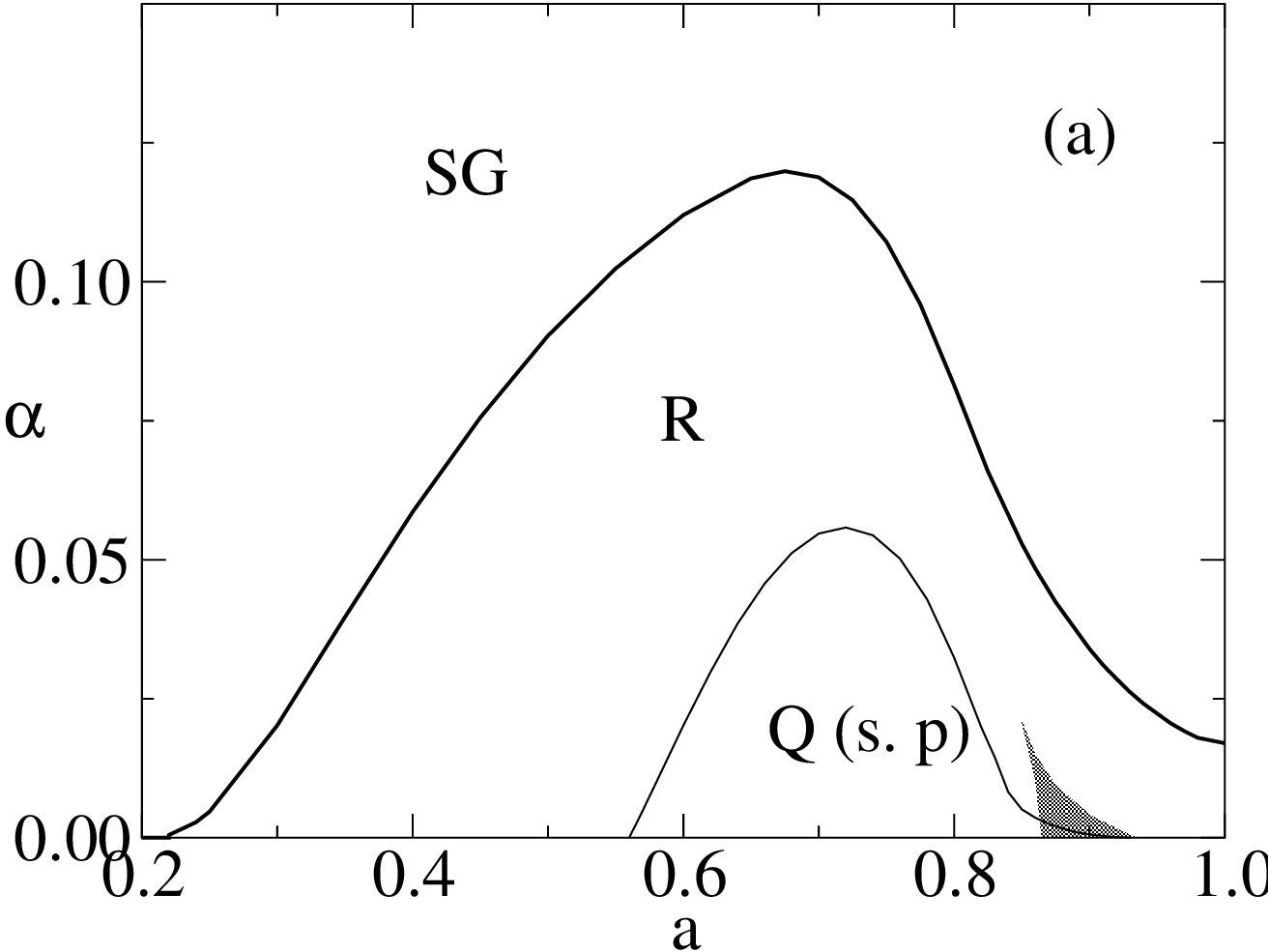}}}
\vspace*{0.5cm}
\center{
\rotatebox{270}{
\includegraphics[width=.35\textwidth,height=7cm]{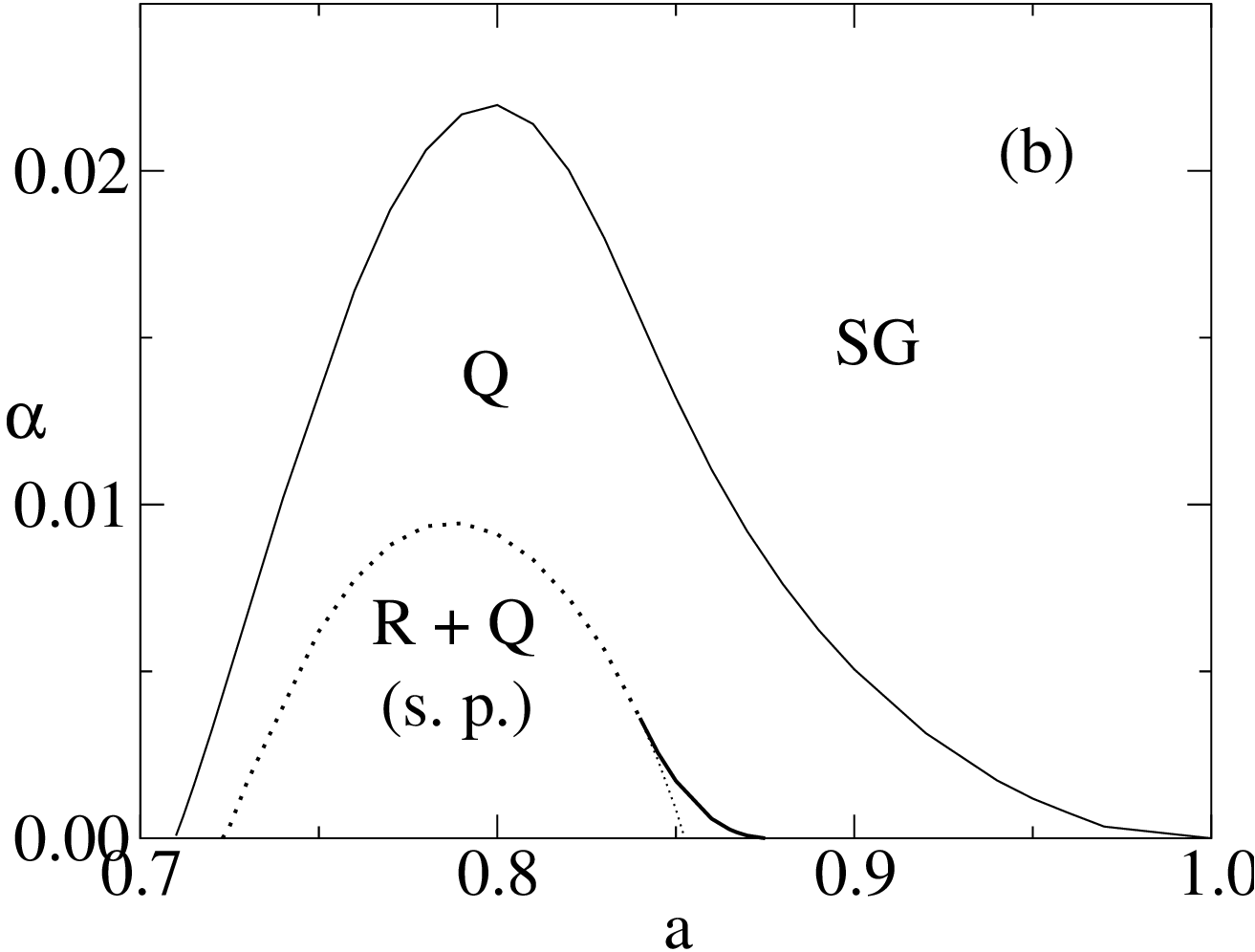}
}}
\caption{The capacity-activity $(\alpha,a)$ phase diagram for the BEG
network at $T=0.4$ (a) and $T=0.8$ (b).
The lines are explained in the text.}
\end{figure}

Thus, as $T$ increases, the useful performance of the network goes
over from the retrieval to the pattern-fluctuation retrieval phase. 
Since the
information content is a common performance measure for both
phases, we show its temperature dependence in an
information-capacity phase diagram in Fig. 5, for $a=0.8$ and
various $T$, where the solid (dotted) lines represent information
due to a stable retrieval (pattern-fluctuation retrieval) phase.

\begin{figure}
\center{
\rotatebox{270}{
\includegraphics[width=.46\textwidth,height=10cm]{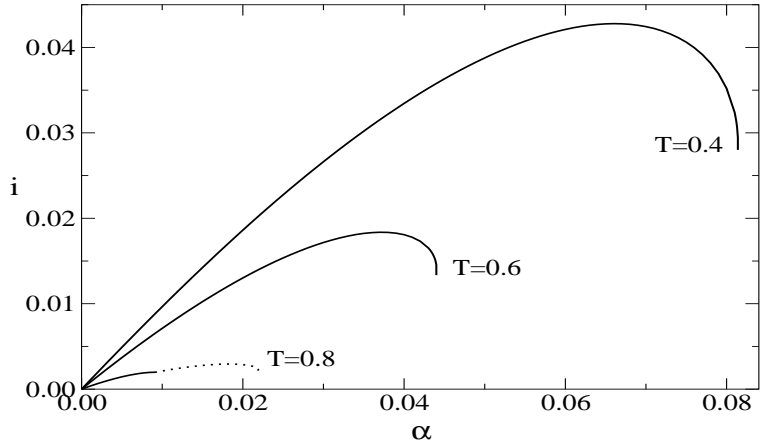}
}}
\caption{The information $i$ as a function of the capacity $\alpha$ for 
several values of $T$ and $a=0.8$.The solid (dashed) line correspond to
the $R$ ($Q$) states.}
\end{figure}

In all the above situations the $SG$ solutions appear as a stable
phase, and we consider now the flow diagrams in the $(l,m)$
order-parameter space. We show the results for $T=0.8$, 
activity $a=0.8$  for $\alpha=0.005$ in the presence of
an $R$  phase in Fig. 6a, and  for $\alpha=0.01$ in the
presence of a $Q$ phase in Fig. 6b. These correspond to 
states on either
side of the dotted phase boundary in Fig. 3b. In the first one
there is a stable retrieval solution and there are two $Q$
saddle points (open circles) and in the second one there is a
stable and an unstable $Q$ solution. In both cases there
is a stable $SG$ solution which can be accessed only below the
lower $Q$ saddle point. The chains of dots actually
indicate time steps and, as can be seen from these figures, the
flows to the stable solutions are considerably delayed by the
saddle points in the form of slow transients of the dynamics. A
remarkable feature of the flow diagrams is the presence of quite
large basins of attraction either to the stable $R$ state or to
the stable $Q$ state, even for the fairly high $T$ (and small
$\alpha$) for this case. Also, not surprisingly, one finds a much
smaller basin of attraction to the $SG$ states.  Similar features have
also been found in the dynamics of the extremely diluted network
except for the $SG$ states, which are absent in that case \cite{BDE02}.

\begin{figure}
\center{
\rotatebox{270}{
\includegraphics[width=.35\textwidth,height=7cm]{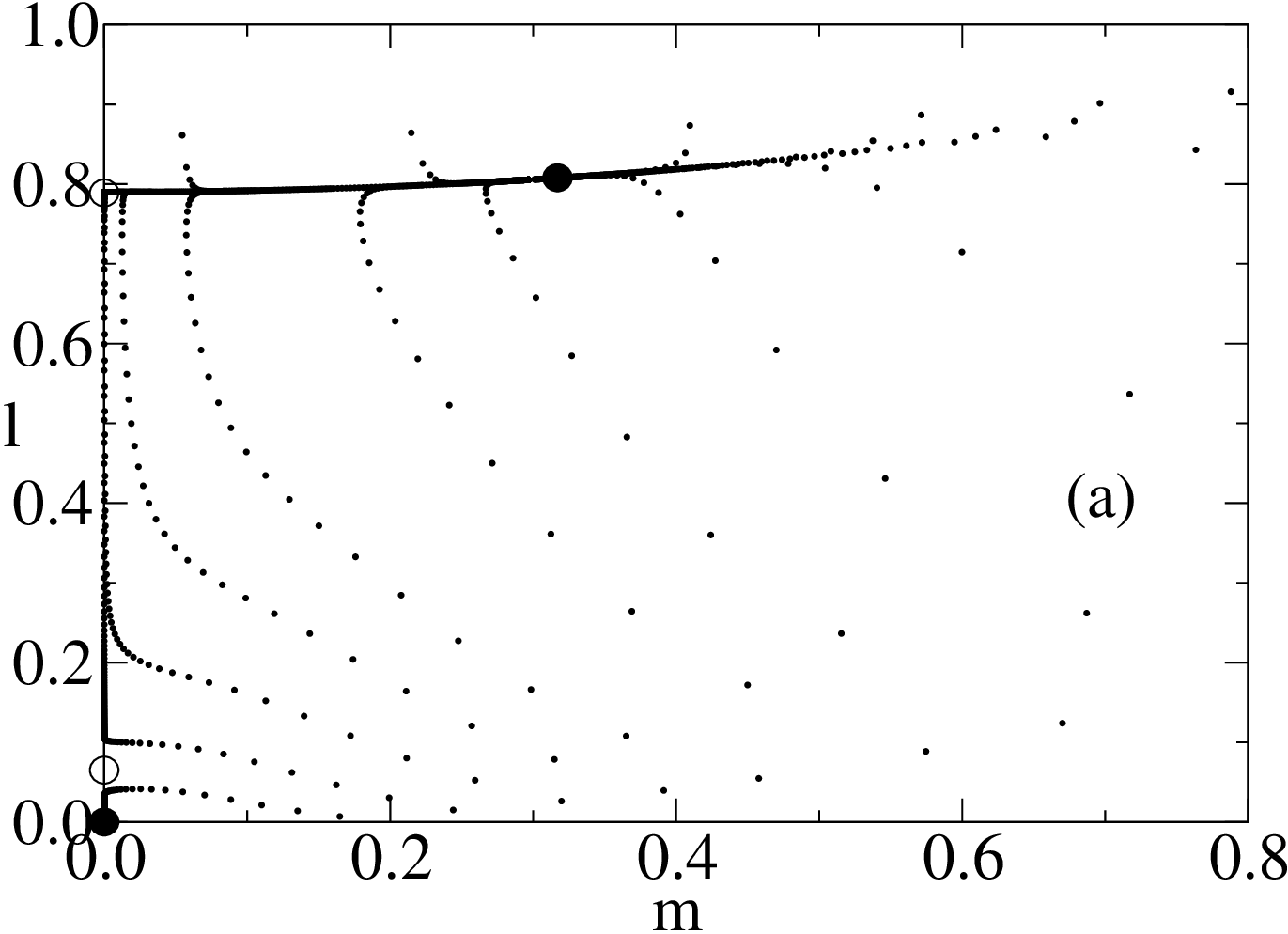}}}
\vspace*{0.5cm}
\center{
\rotatebox{270}{
\includegraphics[width=.35\textwidth,height=7cm]{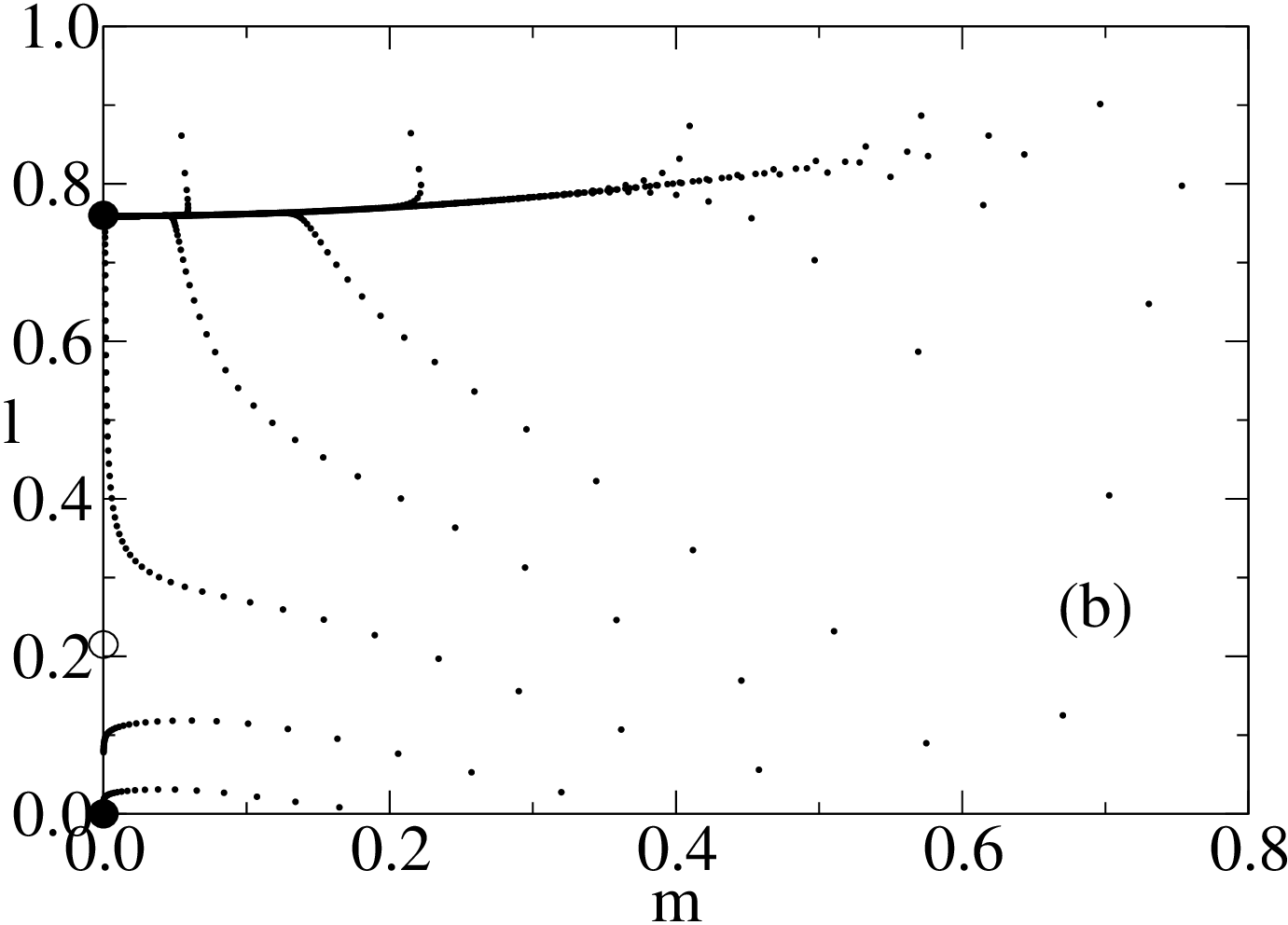}
}}
\caption{Two dimensional retrieval fluctuation overlap $l-m$  flow 
diagrams for the BEG network for
$T=0.8$,  $a=0.8$ and $\alpha=0.005$ (a)  and $\alpha=0.01$ (b).
Open circles are saddle points, closed ones are attractors.}
\end{figure}

We turn now to a brief study of the way in which the phase
diagrams for the layered BEG network turn into those for the
extremely diluted network by means of a variable amplitude $D$ in
the second terms in Eq. (\ref{31}), i.e., 
\begin{equation}
(\Delta^{t+1})^{2}=\alpha A^2q_{0}^{t}
+ D (\chi^{t})^{2}(\Delta^{t})^{2} \,\,, \quad
(\Omega^{t+1})^{2}=\alpha B^{2}q_{0}^{t}
+ D (\psi^{t})^{2}(\Omega^{t})^{2}\,\,\,. 
   \label{D}
\end{equation}
When $D=0$ we have the extremely 
diluted architecture, $D=1$ corresponds to the layered one. To be 
specific, consider the
phase diagram for the layered network at $T=0.8$ shown in Fig. 4b
where the phase boundary for existence of a stable $Q$ phase ends
at $\alpha=0$ when the activity $a=1$. As $D$ decreases from
unity, this part of the phase boundary moves up and the critical
$\alpha$ gradually increases at $a=1$. For $D=0.9$ it becomes
already $\alpha\approx 0.034$. At the same time the maximum
activity $a$ for the lower retrieval phase boundary to a stable
$R$ phase starts to increase towards larger values. The phase boundary
itself 
continues to end at $\alpha=0$. For $D=0.5$, say, the maximum $a$
for retrieval is still less than unity and the $SG$ phase becomes
now restricted to a region on both sides of the left phase
boundary in the $\alpha$ vs. $a$ phase diagram. Ultimately, when
the extremely diluted limit is reached, the stable $R$ phase goes
up to $a=1$, still with $\alpha=0$, with a fairly large $Q$ phase
and the $SG$ phase becomes a self-sustained activity phase $S$,
with $m=0$, $l=0$ and $q_{1}=0$. In part of the phase diagram the
$S$ and $Q$ phases coexist, in accordance with our earlier results
on the extremely diluted network \cite{BDE02}.

\section{Concluding remarks}

We have derived the recursion relations that describe the time
evolution of the macroscopic variables for an exactly solvable
three-state network on a feed-forward layered architecture, optimizing
the mutual information. This so-called layered Blume-Emery-Griffiths 
(BEG) network  shows distinct stationary phase diagrams from either 
its extremely diluted or fully connected versions studied in the
literature. Being a truly dynamical system, there is no 
phase boundary of local stability in the layered network between either
the retrieval, $R$, or the pattern-fluctuation retrieval phase, $Q$,
and the spin-glass phase, $SG$, in contrast to the behavior in the
fully connected network. But this does not mean that
within the retrieval regime the network cannot be trapped in $SG$
states, as it is clear from the flow diagrams. This makes the
layered network different from the extremely diluted
network in which the $R$ or $Q$ states are the only stable states
over most of the regions where these phases exist.

We have found that, in common with both the extremely diluted and the
fully connected network, a stable pattern-fluctuation retrieval phase 
appears only at high
$T$ and for intermediate-to-large, but not full, activity $a$. At
low $T$, in particular at $T=0$, this phase is not stable but is
instead a saddle-point solution. Nevertheless, the BEG layered
network has, selectively, a quite better performance than the 
three-state Ising layered network, not only as far as the retrieval 
capacity is concerned, as in the case of both the fully connected 
and the extremely diluted network, but it also yields a considerably 
larger information content. This additional information is  due 
to the enhancement by the pattern-fluctuation states of the stable
retrieval phase at low and intermediate $T$. The pattern-fluctuation retrieval  
phase, instead, is responsible for a much smaller but non-negligible
information content. 

To summarize, the BEG network on a feed-forward layered structure is 
not only an interesting dynamical system in itself but it also performs
better than other, for instance, Ising layered networks within
relevant regimes of temperature and activity. 

\section*{Acknowledgements}

We thank  D. R. C. Dominguez and T. Verbeiren for critical discussions. 
We are indebted to both the Fund for Scientific Research-Flanders, 
Belgium, and the Funda\c{c}\~ao de Amparo \`a Pesquisa do Estado do Rio 
Grande do Sul (FAPERGS), Brazil, for financial 
support. One of us (DB) thanks the warm hospitality of the Instituto 
de F\'{\i}sica of the UFRGS, Porto Alegre, where this work was initiated.
Both RE and WKT thank the kind hospitality and the support 
of the Institute for Theoretical Physics of the K.U.Leuven, where part 
of the work was done. The work of one of us (WKT) is also partially 
supported by the
Conselho Nacional de Desenvolvimento Cient\'{\i}fico e Tecnol\'ogico
(CNPq), Brazil.


\begin{thebibliography}{}
\bibitem{DK00} D. R. Carreta Dominguez and E. Korutcheva,
Phys. Rev. E {\bf 62}, 2620 (2000).
\bibitem{BV02} D. Boll\'e and T. Verbeiren, Phys. Lett. A {\bf 297}, 156
(2002).
\bibitem{H82} J.J.Hopfield, Proc. Nat. Acad. Sci. USA {\bf 79}, 2554
(1982).
\bibitem{BEG71} M. Blume, V. J. Emery and R. B. Griffiths, Phys. Rev. A {\bf
4}, 1071 (1971); \newline
M. Blume, Phys. Rev. {\bf 141}, 517 (1966); 
\newline
H. W. Capel, Physica {\bf 32}, 966 (1966).
\bibitem{ACN00} J. M. de Ara\'ujo, F. A. da Costa and
F. D. Nobre, Eur. Phys. J. B {\bf 14}, 661 (2000).
\bibitem{DK02} D. R. C. Dominguez, E. Korutcheva, W. K. Theumann, and
R. Erichsen Jr., in Lecture Notes in Computer Science (Springer, Berlin)
{\bf 2415}, 129 (2002).
\bibitem{BDE02} D. Boll\'e, D. R. Dominguez, R. Erichsen, Jr.,
E. Korutcheva and W. K. Theumann, cond-mat/0208281 .
\bibitem{BV03} D. Boll\'e and T. Verbeiren, J. Phys. A {\bf 36}, 295
(2003).
\bibitem{BBS02} D. Boll\'e, J. Busquets Blanco and G. M. Shim, Physica A
{\bf 318}, 613 (2003).
\bibitem{BBSV02} D. Boll\'e, J. Busquets Blanco , G. M. Shim and T.
Verbeiren, cond-mat/0304553
\bibitem{Y89} J. S. Yedidia, J. Phys. A {\bf 22}, 2265 (1989).
\bibitem{BS94} D. Boll\'e, G. M. Shim, B. Vinck, and V. A. Zagrebnov,
J. Stat. Phys. {\bf 74}, 565 (1994).
\bibitem{SNK98} D.L. Schacter, K.A. Norman and W. Koutstaal, Annu. Rev. 
Psychol. {\bf 49}, 289 (1998).
\bibitem{BKS96} E. Barkai, I. Kanter and H. Sompolinsky, Phys. Rev. A
{\bf 41}, 590 (1996).
\bibitem{DKM89} E. Domany, W. Kinzel and R. Meir, J. Phys. A {\bf 22},
2081(1989).
\bibitem{BSV94} D. Boll\'e, G. M. Shim and B. Vinck, J. Stat. Phys.
{\bf 74}, 583 (1994).
\bibitem{BD00} D. Boll\'e and D. Dominguez Carreta, Physica A {\bf 286},
401(2000).
\bibitem{DB98} D. R. C. Dominguez and D. Boll\'e, Phys. Rev. Lett.
{\bf 80}, 2961 (1998).
\bibitem{O96} M. Okada, Neural Netw. {\bf 9}, 1429 (1996).
\bibitem{Sh48} C. E. Shannon,
  Bell Syst. Techn. J. {\bf 27}, 379 (1948).
\bibitem{Bl90} R. E. Blahut,
{\it Principles and Practice of Information Theory}
(Addison-Wesley, Reading, MA 1990), Chapter 5.

\end{thebibliography}
\end{document}